\documentclass[prd,showpacs,floatfix,amsmath,amssymb,nofootinbib,twocolumn]{revtex4-1}
\usepackage{graphicx,color,dcolumn,booktabs,bm}
\usepackage{longtable,comment,braket}
\usepackage{times}
\usepackage{amsmath}
\usepackage{float}
\usepackage{overpic}
\usepackage{amssymb,tipa}
\usepackage{indentfirst}
\usepackage{feynmf}   
\usepackage{slashed}  
\usepackage{cases}
\usepackage{psfrag}
\usepackage{subfigure}
\usepackage{color}
\usepackage{multirow}
\usepackage{lineno}

\usepackage[colorlinks, citecolor=blue,anchorcolor=red,menucolor=red,linkcolor=blue,filecolor=red,runcolor=red,urlcolor=blue,frenchlinks=red]{hyperref}
\usepackage{ulem}
\def\be{\begin{equation}}
\def\ee{\end{equation}}

\begin{document}

\title{Critical behavior and critical exponents of rotating QCD matter}
\author{Kai Xiao$^{a,b}$}
\author{Fei Sun$^{a,b}$} \email{sunfei@ctgu.edu.cn (Corresponding author)}
\author{Shuang Li$^{a,b}$}\email{
lish@ctgu.edu.cn}
\author{Xun Chen$^{c,d}$}

\affiliation{$^a$Department of Physics, China  Three Gorges University, Yichang, 443002, China \\
$^b$Center for Astronomy and Space Sciences, China Three Gorges University, Yichang 443002, China \\
$^c$School of Nuclear Science and Technology, University of South China, Hengyang 421001, China \\
$^d$INFN--Istituto Nazionale di Fisica Nucleare--Sezione di Bari, Via Orabona 4, 70125, Bari, Italy}

\begin{abstract}
We investigate the thermodynamic properties and critical behavior of rotating strongly interacting matter within the two-flavor Nambu–Jona-Lasinio model. The phase structure and critical endpoint (CEP) are determined in the temperature–angular-velocity \((T,\omega)\) plane. Exploiting the thermodynamic conjugacy between the angular velocity \(\omega\) and rotational polarization \(J\), we characterize the CEP through complementary thermal, coexistence, susceptibility, and critical-isotherm responses. The corresponding effective critical exponents are extracted along distinct thermodynamic trajectories and are found to be consistent with \(\alpha_\omega\simeq0\), \(\beta_\omega\simeq1/2\), \(\gamma_\omega\simeq1\), and \(\delta_\omega\simeq3\). These exponents are numerically compatible with the standard scaling relations and form a mutually consistent Landau mean-field scaling pattern. Our results establish a coherent thermodynamic characterization of the rotation-induced CEP and demonstrate how its singular behavior is encoded in rotational response observables associated with the conjugate pair \((\omega,J)\).
\end{abstract}



\maketitle

\section{Introduction\label{sec1}}

The phase structure of strongly interacting matter under extreme conditions is a central topic in quantum chromodynamics (QCD), with profound implications for relativistic heavy-ion collisions, the interiors of neutron stars, and the evolution of the early universe~\cite{Fukushima:2011nu,Alford:2007xm,BraunMunzinger:2016pzz,Bazavov:2017dus}. In particular, non-central heavy-ion collisions generate quark-gluon matter with substantial orbital angular momentum and significant vorticity, providing a natural setting for investigating how rotation influences QCD phase transitions and critical phenomena. 
The large orbital angular momentum carried by the created fireball can be partially transferred to the spin degrees of freedom of its constituents through spin-orbit coupling, resulting in spin polarization of the produced matter, which can manifest itself through the global polarization of emitted hadrons. This phenomenon is often discussed in close analogy with the Barnett effect~\cite{S.J.Barnett,Barnett:1935wyv}. Consequently, quarks and antiquarks may develop preferential spin alignment along the direction of the system's global angular momentum, which is perpendicular to the reaction plane. In a seminal work, Liang and Wang proposed in 2005 that spin-orbit coupling could generate the polarization of strange quarks in non-central heavy-ion collisions~\cite{Liang:2004ph}. Subsequent measurements of global hyperon polarization and vector-meson spin alignment have provided strong evidence linking rotational dynamics with experimentally accessible spin observables~\cite{STAR:2017ckg,STAR:2018gyt,STAR:2020xbm,ALICE:2019aid}. These developments have stimulated extensive lattice QCD studies of rotating QCD matter~\cite{Braguta:2020biu,Braguta:2021jgn,Braguta:2022str,Chernodub:2022veq,Braguta:2023yjn,Braguta:2023iyx,Yang:2023vsw,Braguta:2024zpi,Braguta:2025yud,Ding:2026gtv} and motivated further theoretical investigations.

Rotating QCD matter has attracted considerable attention due to its rich phase structure, including phenomena such as pion superfluidity~\cite{Liu:2017spl,Cao:2019ctl,Chen:2019tcp}, $\rho$-meson superconductivity~\cite{Zhang:2018ome,Cao:2020pmm}, and rotation-induced modifications of chiral and deconfinement phase transitions~\cite{Chen:2015hfc,Jiang:2016wvv,Ebihara:2016fwa,Chernodub:2016kxh,Chernodub:2017ref,Wang:2018sur,Sun:2021hxo,Xu:2022hql,Sun:2023kuu,Chen:2020ath,Chen:2023cjt,Braga:2023qej,Ambrus:2023bid,Sun:2024anu,Wang:2024szr,Hua:2024bwn,Zhao:2022uxc,Golubtsova:2022ldm,Chen:2022mhf,Chen:2022smf,Yadav:2022qcl,Braga:2022yfe,Cartwright:2021qpp,Golubtsova:2021agl,Chernodub:2020qah,Fujimoto:2021xix,Du:2026hcr}. The angular velocity has therefore emerged as an important control parameter in the study of the QCD phase diagram. Since it couples directly to angular momentum and spin degrees of freedom, it can significantly influence the thermodynamic properties~\cite{Wang:2018sur,Sun:2023yux} and phase-transition behavior of strongly interacting matter.

Among the features of the QCD phase diagram, the critical endpoint (CEP) plays a particularly important role, as it represents the termination of the first-order phase-transition line and the emergence of the crossover region. Near the CEP, critical fluctuations, long-range correlations, and universal scaling behavior become prominent. These singular properties are characterized by the critical exponents $\alpha$, $\beta$, $\gamma$, and $\delta$, which provide key signatures of the universality class of the phase transition. Critical phenomena and related critical exponents have been extensively investigated within various theoretical frameworks, including the Nambu--Jona-Lasinio (NJL) model, quark-meson (QM) model, holographic AdS/QCD approaches, Dyson-Schwinger equations (DSE), and the functional renormalization group (FRG) method~\cite{Costa:2008yh,DeWolfe:2010he,Du:2015psa,Chen:2018msc,Chen:2018vty,Fu:2019hdw,Gao:2020fbl,Gunkel:2021oya,Chen:2023cjt,Zhao:2023gur,Fu:2024wkn,Chen:2024hki,Cai:2024eqa,Sun:2025uga,Li:2026lxx,Fu:2026qnl}. Experimentally, the RHIC Beam Energy Scan (BES) program has been designed to search for the CEP through measurements of higher-order fluctuations of conserved charges~\cite{STAR:2020tga}. Meanwhile, recent lattice QCD studies have explored the CEP using several approaches, including Yang-Lee edge singularities
and contours of constant entropy density ~\cite{Basar:2023nkp,Clarke:2024ugt,Shah:2024img,Borsanyi:2025dyp,Adam:2025phc}. These efforts have significantly improved our understanding of the CEP and its associated critical behavior. However, in realistic heavy-ion collisions, QCD matter is also subject to extremely strong vorticity, indicating that angular velocity may serve as an additional control parameter for the QCD phase structure beyond temperature and baryon chemical potential. Compared with the conventional $(T,\mu_B)$ phase diagram, the rotational effects on critical phenomena remain largely unexplored. Despite extensive studies of rotation-induced modifications of the QCD phase structure, a systematic characterization of the critical scaling near a rotation-induced CEP remains lacking. In particular, it is unclear whether complementary thermal and rotational response observables exhibit a mutually consistent scaling structure.

In this work, we employ the NJL model~\cite{Nambu:1961tp,Nambu:1961fr,Klevansky:1992qe,Hatsuda:1994pi,Buballa:2003qv,Jiang:2016wvv}, which effectively describes the essential low-energy features of QCD, especially chiral symmetry breaking and restoration, to investigate rotating QCD matter. We calculate the thermodynamic potential, determine the phase structure in the $(T,\omega)$ plane, locate the CEP, analyze the associated critical behavior,  and establish a thermodynamic scaling framework for the rotation-induced CEP, whose internal consistency is tested through four complementary response observables. 

This paper is organized as follows. In Sec.~\ref{sec2}, we introduce the two-flavor NJL model in a rotating frame, define the effective critical exponents in the \((T,\omega)\) plane, and derive the corresponding thermodynamic observables. In Sec.~\ref{sec3}, we analyze the phase structure, investigate the critical behavior of thermodynamic observables near the CEP, and extract the effective critical exponents from the numerical results. Sec.~\ref{sec4} presents our conclusions and discusses possible directions for future studies.

\section{FORMALISM}
\label{sec2}

To describe matter under global rigid rotation, it is convenient to adopt a co-rotating reference frame, in which the effects of rotation can be incorporated through the corresponding spacetime geometry. In this framework, the metric tensor describing the rotating system is given by
\begin{eqnarray}
g_{\mu\nu}=
\left(
\begin{array}{cccc}
1-\vec{v}^{\,2} & -v_1 & -v_2 & -v_3 \\
-v_1 & -1 & 0 & 0 \\
-v_2 & 0 & -1 & 0 \\
-v_3 & 0 & 0 & -1
\end{array}
\right),
\end{eqnarray}
where \(v_i\) denotes the local velocity induced by rotation. We consider a uniformly rotating system around the \(z\)-axis with angular velocity
\(\vec{\omega}=\omega \hat{z}\).

In a rotating frame, the standard free Dirac Lagrangian in flat Minkowski spacetime must be modified to account for the nontrivial spacetime structure. The effects of rotation are introduced through the spin connection \(\Gamma_\mu\). Specifically, two geometric modifications are required when generalizing the Dirac equation from flat to curved spacetime. First, the constant gamma matrices \(\gamma^\mu\) are replaced by spacetime-dependent matrices \(\bar{\gamma}^{\mu}(x)\), which satisfy the Clifford algebra defined by the curved metric \(g_{\mu\nu}(x)\). Second, the ordinary derivatives are replaced by spinor covariant derivatives,
\(
D_\mu=\partial_\mu+\Gamma_\mu,
\)
where \(\Gamma_\mu\) represents the spin connection.
Keeping only terms up to linear order in the angular velocity, the two-flavor NJL model in a rotating frame can be written as~\cite{Jiang:2016wvv}
\begin{equation}
\begin{aligned}
\mathcal{L} =
&\, \bar{\psi}\left(i\gamma^\mu\partial_\mu-\hat m+\gamma^0\mu\right)\psi\\
&+\bar\psi\left[(\bm{\omega}\times\bm{x})\cdot(-i\bm{\nabla})+\bm{\omega}\cdot\bm{S}\right]\psi\\
&+G\left[
(\bar{\psi}\psi)^2+
(\bar{\psi}i\gamma^5\bm{\tau}\psi)^2
\right].
\label{eq:Lagrangian}
\end{aligned}
\end{equation}
Here, \(\hat m\) denotes the current quark mass matrix in flavor space. In the isospin-symmetric limit adopted here, the \(u\) and \(d\) quark masses are taken to be equal, \(m_u=m_d=m\), such that \(\hat m=m\mathbf{1}_f\), \(G\) is the four-fermion coupling constant, \(\bm{S}\) is the spin operator, and \(\bm{\tau}\) represents the Pauli matrices in flavor space. The rotational contribution in Eq.~\eqref{eq:Lagrangian} couples the angular velocity to both orbital and spin angular momenta of quarks. The scalar and pseudoscalar interaction channels are retained to preserve the chiral symmetry structure of the two-flavor NJL model. Within the mean-field approximation, we consider a finite scalar condensate,
\(
\langle\bar{\psi}\psi\rangle\neq0,
\)
while the pseudoscalar condensate is assumed to vanish,
\(
\langle\bar{\psi}i\gamma^5\vec{\tau}\psi\rangle=0.
\)

Starting from the finite-temperature partition function in the imaginary-time formalism
\begin{eqnarray}
{\cal Z}=\int D[\bar{\psi}]D[\psi]e^{-S_E},
\end{eqnarray}
where \(S_E\) is the Euclidean action obtained after Wick rotation, the thermodynamic potential can be derived using the standard finite-temperature field-theory formalism. The volume-averaged grand-potential density is defined as
\(
\Omega_{averaged}=-\frac{T}{V}\ln{\cal Z}.
\)
For details of the derivation, we refer the reader to Refs.~\cite{Jiang:2016wvv,Sun:2021hxo}. Because global rotation breaks translational invariance in the transverse direction, the thermodynamic quantities generally depend on the distance from the rotation axis. We therefore introduce a local grand-potential density. At vanishing quark chemical potential, it is given by
\begin{equation}
\begin{aligned}
\Omega=&
\frac{(M-m)^2}{4G}
-\frac{N_fN_c}{2\pi^2}
\sum_n
\int_0^\Lambda k_t dk_t
\int_{-\sqrt{\Lambda^2-k_t^2}}^{\sqrt{\Lambda^2-k_t^2}}dk_z
\\
&\times
\mathcal{J}_n(k_t r)
T\ln
\left[
2\cosh
\left(
\frac{\varepsilon_n}{2T}
\right)
\right],
\label{eq:Omega}
\end{aligned}
\end{equation}
where \(N_c=3\) and \(N_f=2\) denote the numbers of colors and flavors, respectively. The parameter \(\Lambda\) represents the three-momentum cutoff, and \(n\in\mathbb{Z}\) labels the quantum number for the \(z\)-component of angular momentum.
The quasiparticle energy modified by rotation is given by
\(\varepsilon_n=E_k-
\left(n+\frac12\right)\omega,\)
where
\(
E_k=\sqrt{k_z^2+k_t^2+M^2}.
\)
The constituent quark mass is determined by
\(
M=m-2G\langle\bar{\psi}\psi\rangle.
\)
For convenience, we define
\(
\mathcal{J}_n(k_t r)
=
J_n^2(k_t r)+J_{n+1}^2(k_t r).
\)

The equilibrium state is obtained from the stationary condition
\begin{equation}
\frac{\partial\Omega}{\partial M}=0,
\qquad
\frac{\partial^2\Omega}{\partial M^2}>0,
\label{eq:stationary}
\end{equation}
where the second condition guarantees local thermodynamic stability. When multiple locally stable solutions exist, the physical equilibrium state is selected by the global minimum of \(\Omega\). Away from the CEP, local thermodynamic stability requires \(\partial^2\Omega/\partial M^2>0\). At the CEP, the curvature in the critical direction vanishes. In addition, for a rotating cylindrical system, causality requires that the local tangential velocity remain below the speed of light. 
Solving the stationary condition yields the gap equation
\begin{equation}
\begin{aligned}
M=&m+
G\frac{N_cN_f}{2\pi^2}
\sum_{n=-\infty}^{\infty}
\int_0^\Lambda k_tdk_t
\int_{-\sqrt{\Lambda^2-k_t^2}}^{\sqrt{\Lambda^2-k_t^2}}dk_z
\\
&\times
\mathcal{J}_n(k_t r)
\frac{M}{E_k}
(f_- - f_+),
\label{eq:gap}
\end{aligned}
\end{equation}
where the Fermi distribution functions are defined as
\begin{equation}
f_+
=
\frac{1}{1+e^{\varepsilon_n/T}},
\qquad
f_-
=
\frac{1}{1+e^{-\varepsilon_n/T}}.
\end{equation}

For a rotating system, in terms of local thermodynamic densities, the first law takes the form
\begin{equation}
d\varepsilon = T\,ds + \mu\,dn + \omega\,dJ,
\end{equation}
where \(\varepsilon\),  \(s\), \(n\), and \(J\) denote the energy density, entropy density, quark number density, and rotational polarization (angular momentum density), respectively. The corresponding grand potential density is obtained through the Legendre transformation \cite{Landau_1980,Chernodub:2016kxh,Zhao:2022uxc}
\begin{equation}
\Omega = \varepsilon - Ts - \mu n - \omega J.
\label{potentialdensity}
\end{equation}
Combining the first law relation, its differential is therefore
\begin{equation}
d\Omega = -s\,dT - n\,d\mu - J\,d\omega.
\end{equation}
Treating \(\Omega\) as a function of \(T\), \(\mu\), and \(\omega\), comparison with its total differential immediately gives
\begin{equation}
s = -\left(\frac{\partial \Omega}{\partial T}\right)_{\mu,\omega},
n = -\left(\frac{\partial \Omega}{\partial \mu}\right)_{T,\omega},
J = -\left(\frac{\partial \Omega}{\partial \omega}\right)_{T,\mu}.
\end{equation}

We next introduce the effective critical exponent \(\alpha_\omega\), which characterizes the singular behavior of the specific heat density near the CEP. 
The subscript \(\omega\) indicates that these effective exponents are defined for the rotating system in the (\(T,\omega)\) parameter space.
As the temperature approaches the CEP, the specific heat density follows the scaling relation
\begin{equation}
C_J \sim |t|^{-\alpha_\omega},
\end{equation}
where \(t=(T-T_{\mathrm{CEP}})/T_{\mathrm{CEP}}\) denotes the reduced temperature. The specific heat density at fixed rotational polarization is defined by
\begin{equation}
C_J=
T\left(\frac{\partial s}{\partial T}\right)_J=
-T
\left(
\frac{\partial^2 \bar{\Omega}}{\partial T^2}
-\left(
\frac{\partial^2 \bar{\Omega}}{\partial T \partial \omega}
\right)^2 \left(\frac{\partial^2 \bar{\Omega}}{\partial \omega^2}\right)^{-1}\right).
\end{equation}
Here, \(\bar{\Omega}\) denotes the equilibrium grand-potential density obtained after solving the gap equation for the constituent quark mass. Consequently, the derivatives entering \(C_J\) are taken with respect to the equilibrium-reduced thermodynamic potential and implicitly include the self-consistent response of \(M\).

We next consider the critical exponent \(\beta_\omega\), which describes the scaling behavior of an order-parameter-like quantity near the CEP. In the rotating system, the thermodynamic quantity conjugate to the angular velocity is the rotational polarization \(J\). 
Although \(J\) is not the conventional chiral order parameter, its singular part can couple to the critical mode through the mixed \(M-\omega\) dependence of the thermodynamic potential. Provided that \(\Omega_{\omega M}^{\rm CEP}\neq0\), \(J\) inherits the leading singular scaling associated with the soft chiral mode.
The rotational polarization is explicitly given by
\begin{equation}
\begin{aligned}
J
=&
\frac{N_cN_f}{4\pi^2}
\sum_n
\int k_tdk_t
\int dk_z\,
\mathcal{J}_n(k_tr)
\\
&\times
\left(n+\frac12\right)
(f_+-f_-).
\label{eq:J-explicit}
\end{aligned}
\end{equation}
Across the first-order phase transition line, \(J\) exhibits a discontinuity due to the coexistence of two thermodynamic phases. Therefore, the difference between the two branches is defined as
\begin{equation}
\Delta J
=
J_>(T,\omega,M_>)
-
J_<(T,\omega,M_<),
\end{equation}
and its critical scaling behavior is characterized by
\begin{equation}
|\Delta J|\sim |t|^{\beta_\omega}.
\end{equation}

The exponent \(\gamma_\omega\) describes the divergence of the rotational susceptibility near the CEP. Since the constituent quark mass is determined self-consistently through the gap equation, the susceptibility should be evaluated as a total derivative along the equilibrium trajectory,
\begin{equation}
\chi_\omega
=
\left(\frac{d J}{d\omega}\right)_T.
\end{equation}
Applying the chain rule together with the stationarity condition gives
\begin{equation}
\chi_\omega
=
\left.
\frac{\partial J}{\partial\omega}
\right|_{M,T}
+
\left.
\frac{\partial J}{\partial M}
\right|_{\omega,T}
\left.
\frac{\partial M}{\partial\omega}
\right|_T
=
-\frac{\partial^2\Omega}{\partial\omega^2}
+
\frac{
\left(
\frac{\partial^2\Omega}{\partial\omega\partial M}
\right)^2
}{
\frac{\partial^2\Omega}{\partial M^2}
}.
\label{eq:chi-definition}
\end{equation}
The explicit form of the rotational susceptibility is
\begin{equation}
\begin{aligned}
\chi_\omega
=&
\frac{N_cN_f}{2\pi^2T}
\sum_n
\int k_tdk_t
\int dk_z\,
\mathcal{J}_n(k_tr)
\left(n+\frac12\right)^2
f_+f_-
\\
&+
\left(
\frac{N_cN_f}{2\pi^2T}
\sum_n
\int k_tdk_t
\int dk_z\,
\mathcal{J}_n(k_tr)
\frac{(n+\frac12)M}{E_k}
f_+f_-
\right)^2
\\
&\times
\left(
\frac{1}{2G}-B
\right)^{-1},
\end{aligned}
\label{eq:chi}
\end{equation}
where the first term corresponds to the direct response of the rotating system to angular velocity, while the second term represents the contribution arising from the coupling of the rotational response to the soft chiral mode through the self-consistent constituent mass. The function \(B\) is defined as
\begin{equation}
\begin{aligned}
B
=&
\frac{N_cN_f}{2\pi^2}
\sum_n
\int k_tdk_t
\int dk_z\,
\mathcal{J}_n(k_tr)
\\
&\times
\left[
\frac{k_t^2+k_z^2}{2E_k^3}
(f_--f_+)
+
\frac{M^2}{TE_k^2}
f_+f_-
\right].
\end{aligned}
\end{equation}
When the system approaches the CEP along a trajectory tangent to the phase boundary, the rotational susceptibility obeys the scaling law
\begin{equation}
\chi_\omega\sim |t|^{-\gamma_{\omega}},
\end{equation}
from which the effective critical exponent \(\gamma_\omega\) can be extracted. Along this tangent trajectory, \(t\) is used as the path parameter.

We finally consider the effective critical-isotherm exponent
\(\delta_\omega\), which characterizes the nonlinear rotational response
of the system at the CEP. In the rotating system, the angular velocity
\(\omega\) acts as a thermodynamic control parameter conjugate to the
rotational polarization (see Eq. \eqref{potentialdensity}).
This thermodynamic conjugacy is analogous to that between the chemical
potential and the quark-number density,
\(n=-\partial\Omega/\partial\mu\), in the conventional \(T\!-\!\mu\)
phase diagram.
Although \(J\) is not the conventional chiral order parameter, it contains a singular contribution associated with the critical mode. Indeed, near the CEP, along the critical isotherm, the rotational polarization can be expanded as
\begin{equation}
J-J_{\rm CEP}
\simeq
-\Omega_{\omega M}^{\rm CEP}(M-M_{\rm CEP})
-\Omega_{\omega\omega}^{\rm CEP}(\omega-\omega_{\rm CEP})
+\cdots .
\end{equation}
For \(\Omega_{\omega M}^{\rm CEP}\neq0\), the first term couples the rotational response directly to the critical mode associated with  \(M\), so that the singular part of \(J\) inherits its leading critical scaling. We therefore characterize the critical-isotherm
response in terms of the reduced quantities
\begin{equation}
\widetilde{J}=(J-J_{\rm CEP})/J_{\rm CEP},
\qquad
\widetilde{\omega}=(\omega-\omega_{\rm CEP})/\omega_{\rm CEP},
\end{equation}
and define the effective exponent \(\delta_\omega\) through
\begin{equation}
|\widetilde{J}|
\sim
|\widetilde{\omega}|^{1/\delta_{\omega}},
\qquad
T=T_{\rm CEP}.
\end{equation}
Here, \(\delta_{\omega}\) is obtained from \(\frac1{\delta_{\omega}} = \frac{\Delta\ln|\widetilde J|}{\Delta\ln|\widetilde\omega|}\).
Physically, this scaling relation describes how the rotational response
approaches its critical value as the angular velocity is varied toward
\(\omega_{\rm CEP}\) along the critical isotherm. Thus,
\(\delta_{\omega}\) provides a measure of the nonlinear rotational response
associated with the CEP.  It should be pointed out that the reduced quantities \(t\) and \(\widetilde{\omega}\) need not coincide exactly with the thermal- and field-like scaling fields of the asymptotic critical theory; the exponents obtained here are therefore understood as effective exponents associated with the specified thermodynamic trajectories.

\section{Results and discussion}
\label{sec3}

In this section, we present our numerical results for the constituent quark mass, the chiral phase structure, and the critical behavior of rotating QCD matter within the two-flavor NJL model. We adopt the parameter set from Ref.~\cite{Kohyama:2016fif}, with 
\(
m=0.005~\mathrm{GeV},
G=3.672~\mathrm{GeV}^{-2}\), and 
\(\Lambda=0.6816~\mathrm{GeV}
\).
Throughout this section, the radial coordinate is fixed at \(r=0.1~\mathrm{GeV}^{-1}\), following Ref. \cite{Jiang:2016wvv}. At the fixed radial position used in the present local analysis, the causality condition \(\omega r<1\) is safely satisfied due to \(\omega_{\rm CEP}r  \simeq0.0644\ll1\).
Here, the present calculation should be interpreted as a local thermodynamic analysis at a fixed radial position rather than as a fully volume-averaged description of a finite rotating cylinder. The angular-momentum sum is truncated to \(-5\leq n\leq5\), as in Ref. \cite{Sun:2021hxo}. Although the sum formally extends over all integer \(n\), the series converges rapidly and this truncation is sufficient for the numerical accuracy required here.

To characterize the chiral phase transition in rotating QCD matter, we analyze the thermodynamic potential as a function of the constituent quark mass \(M\). Within the mean-field approximation, \(M\) is directly related to the chiral condensate and serves as an effective order parameter for chiral symmetry breaking. Therefore, the structure of \(\Omega(M)\) provides direct information on the phase nature and the evolution of the chiral transition. The equilibrium state is determined by the stationary condition
\(
\frac{\partial\Omega}{\partial M}=0,
\)
and the extrema of \(\Omega(M)\) correspond to possible equilibrium or metastable states. Multiple local minima may signal metastability, while two degenerate minima at the transition point indicate phase coexistence characteristic of a first-order transition. By contrast, a single smooth minimum is associated with crossover behavior.

\begin{figure}[htbp]
\centering
\subfigure{\includegraphics[width=0.48\textwidth]{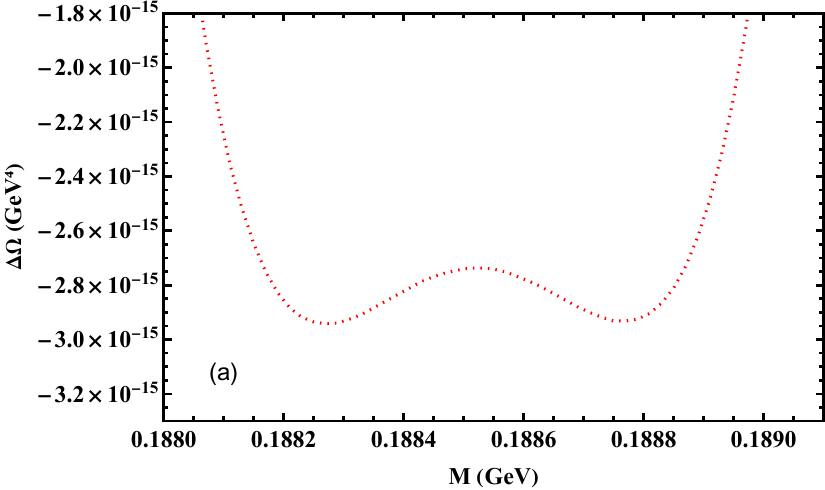}}
\vspace{0.0cm}
\subfigure{\includegraphics[width=0.48\textwidth]{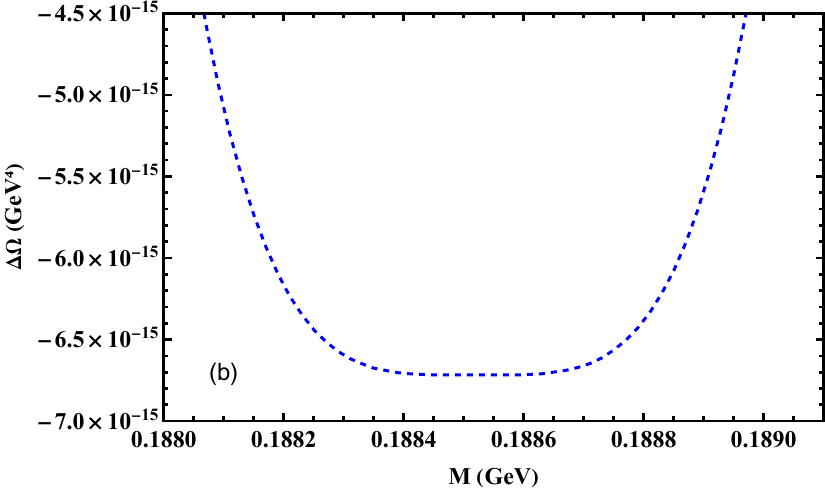}}
\vspace{0.0cm}
\subfigure{\includegraphics[width=0.48\textwidth]{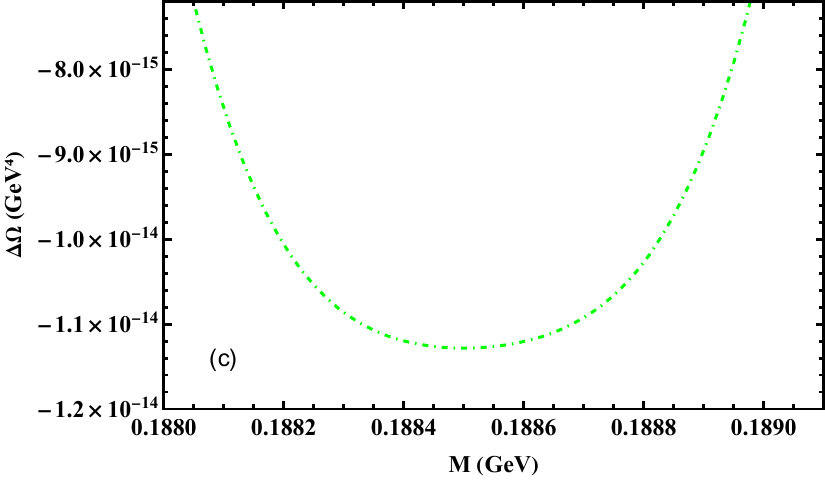}}
\vspace{0.0cm}
\caption[]{(Color online) Shifted grand-potential density \(\Delta\Omega\) as a function of the constituent quark mass \(M\) at representative points in the \(T\!-\!\omega\) phase diagram: (a) the first-order transition line, (b) the CEP, and (c) the crossover region. Here, \(\Delta\Omega=\Omega-\Omega_{\rm ref}\), with \(\Omega_{\rm ref}\) a constant chosen separately for each panel. The shift only changes the reference of the potential and leaves the extrema and phase structure unchanged. The evolution from a double-well potential with two degenerate minima, to a nearly flat potential at the CEP, and finally to a single-minimum structure illustrates the change of the chiral phase transition across the CEP.
}
\label{fig:Omega-M}
\end{figure}

Fig.~\ref{fig:Omega-M} illustrates the evolution of the grand-potential landscape from the first-order coexistence region, through the CEP, to the crossover region. Along the first-order transition line, \(\Delta\Omega(M)\) exhibits two degenerate minima separated by a finite barrier, corresponding to two coexisting phases with different constituent quark masses. The resulting discontinuity of the order parameter is the characteristic signature of a first-order transition. At the CEP, the two minima merge and the potential becomes significantly flattened in the chiral direction. This critical behavior is characterized by the vanishing curvature,
\(
\frac{\partial^2\Omega}{\partial M^2}\rightarrow0,
\)
which signals the softening of the chiral mode and the associated enhancement of thermodynamic response. In contrast, the crossover region contains only a single smooth minimum, corresponding to a continuous evolution of the order parameter without phase coexistence. The CEP is determined by tracking the evolution of the thermodynamic potential from the first-order region toward the crossover region and identifying the point where the discontinuity of the order parameter vanishes. Specifically, we define
\(
\Delta M=M_{\mathrm{broken}}-M_{\mathrm{restored}},
\)
where \(M_{\mathrm{broken}}\) and \(M_{\mathrm{restored}}\) correspond to the two coexisting minima along the first-order boundary. As the CEP is approached,
\(
\Delta M\rightarrow0,
\)
and the CEP is numerically located at \(
T_{\mathrm{CEP}}\simeq0.0202339~\mathrm{GeV}, 
\omega_{\mathrm{CEP}}\simeq0.6440127~\mathrm{GeV}
\), where the coordinates are rounded to a reasonable number of significant digits for presentation. Higher-precision numerical values are retained internally throughout the near-critical analysis to ensure the numerical stability of the extracted effective critical exponents.

\begin{figure}[htbp]
\centering
\includegraphics[width=0.48\textwidth]{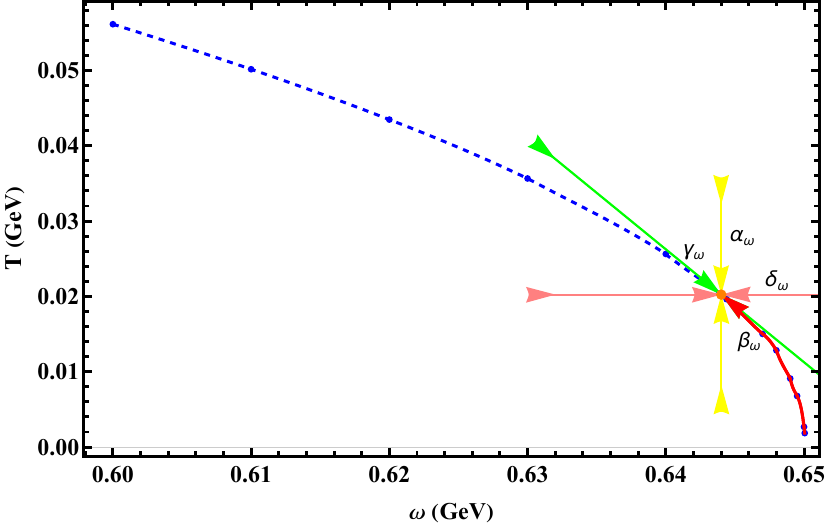}
\caption{(Color online) Phase diagram in the \(T-\omega\) plane with schematic indications of the trajectories used to extract the effective critical exponents. The crossover line (which is defined by the maximum of \(-\partial M/\partial T\) at fixed \(\omega\)) is plotted as the blue dashed curve, while the red solid curve corresponds to the first-order phase boundary. The orange filled circle indicates the location of the CEP. The arrows schematically indicate the directions of the thermodynamic trajectories used in the scaling analysis.}
\label{exponents_directions.pdf}
\end{figure}

Fig.~\ref{exponents_directions.pdf} summarizes the phase structure of the rotating system in the \(T-\omega\) plane together with the thermodynamic trajectories employed in the critical-scaling analysis. The phase boundaries are obtained numerically, whereas the arrows schematically indicate the trajectories used to extract the effective critical exponents. The first-order transition line terminates at the CEP, beyond which the system undergoes a smooth crossover. Here, the (pseudocritical) crossover line is defined by the maximum of \(-\partial M/\partial T\) at fixed angular velocity \(\omega\). To characterize the critical behavior from complementary thermodynamic directions, the effective exponent \(\alpha_\omega\) is extracted by approaching the CEP along the constant-angular-velocity trajectory \(\omega=\omega_{\rm CEP}\), \(\beta_\omega\) is determined along the first-order coexistence line, \(\gamma_\omega\) is obtained along a trajectory tangent to the phase boundary at the CEP (The tangent line is constructed as \(T-T_{\rm CEP} = k (\omega-\omega_{\rm CEP}),\)
where the slope \(k=-0.1499\) is numerically evaluated from the phase boundary near the CEP), and \(\delta_\omega\) is evaluated along the critical isotherm \(T=T_{\rm CEP}\). These distinct trajectories provide a unified framework for probing different aspects of the singular thermodynamic response near the rotation-induced CEP.

\begin{figure}[htbp]
\centering
\includegraphics[width=0.48\textwidth]{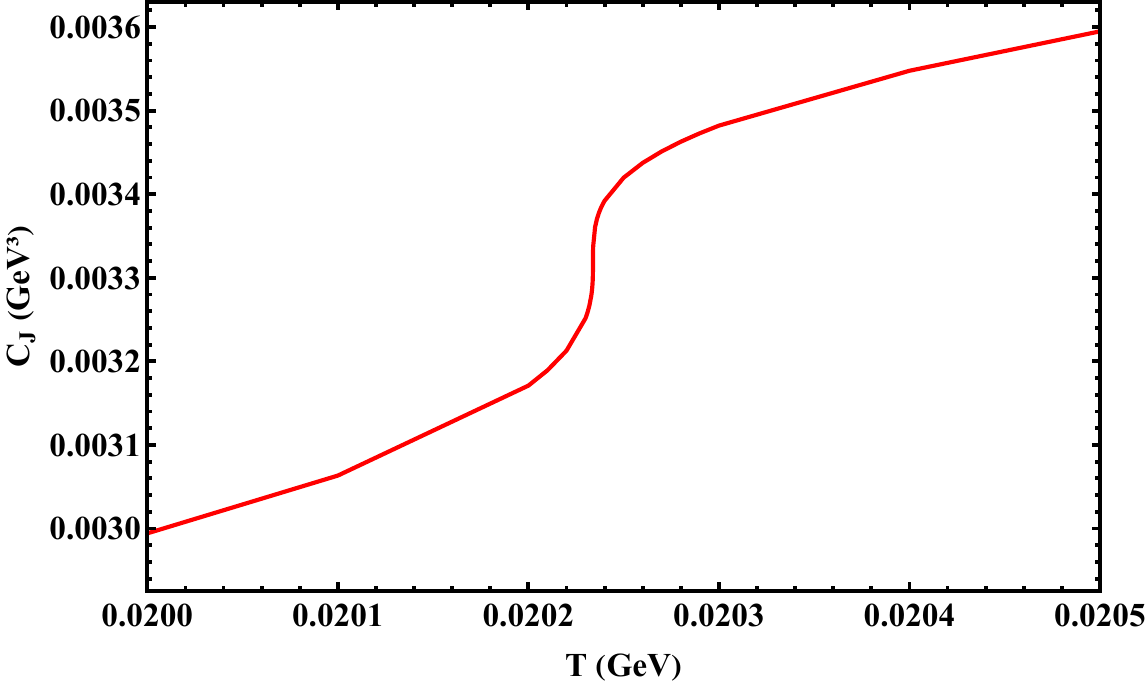}
\caption{(Color online) Specific heat density \(C_J\) as a function of temperature at \(\omega=\omega_{\mathrm{CEP}}\). }
\label{C_vs_T.pdf}
\end{figure}

To characterize the singular behavior of the thermodynamic system near the CEP, we first investigate the effective critical exponent \(\alpha_\omega\) associated with the specific heat density. To investigate its critical behavior, we evaluate \(C_J\) along the trajectory \(\omega=\omega_{\rm CEP}\) and approach the CEP by varying the temperature. Here, the fixed-\(J\) condition specifies the thermodynamic constraint defining the response function \(C_J\), whereas \(\omega=\omega_{\rm CEP}\) specifies the trajectory along which this response is probed near the CEP.
The specific heat density characterizes the thermal response of the system and therefore provides a useful probe of critical behavior. As shown in Fig.~\ref{C_vs_T.pdf}, the specific heat density exhibits a pronounced enhancement when the system approaches the CEP.

\begin{figure}[htbp]
\centering
\includegraphics[width=0.48\textwidth]{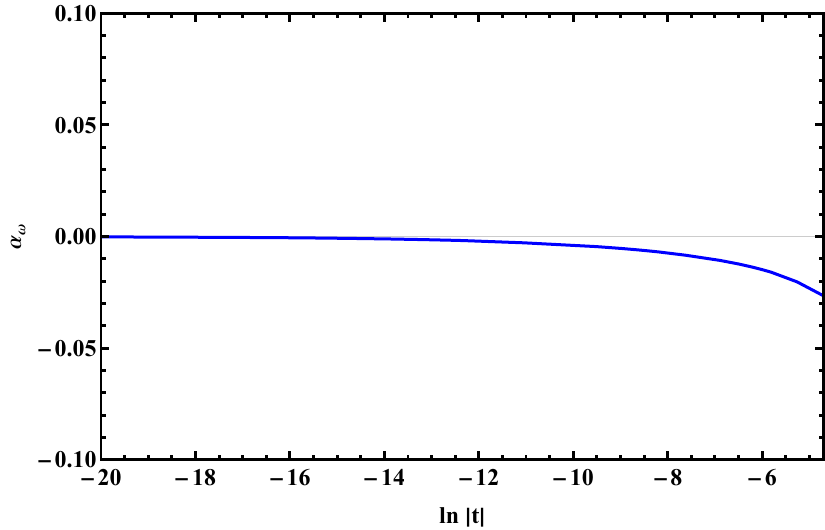}
\caption{(Color online) Effective critical exponent \(\alpha_\omega\) as a function of \(\ln |t|\) along \(\omega=\omega_{\mathrm{CEP}}\).}
\label{alpha_vs_lnt.pdf}
\end{figure}

The effective exponent \(\alpha_\omega\) is extracted from the local scaling behavior of the specific heat density near the CEP. Specifically, it is obtained from the logarithmic slope of \(C_J\) with respect to the reduced temperature in the \(\ln C_J-\ln|t|\) representation using adjacent numerical data points. This local-slope procedure avoids the need to select a global fitting window and allows the evolution of the effective exponent to be monitored as the CEP is approached.
\[
\alpha_\omega(t_i)
=
-\frac{\Delta\ln C_J}{\Delta\ln|t|}
=
-\frac{\ln C_J(t_{i+1})-\ln C_J(t_i)}
{\ln|t_{i+1}|-\ln|t_i|},
\]
where adjacent numerical data points are used to evaluate the finite difference. 
The remaining effective critical exponents \(\beta_\omega, \gamma_\omega\), and \(\delta_\omega\) are extracted in an analogous fashion via adjacent-point logarithmic slopes, following the same numerical procedure described for \(\alpha_\omega\).
The resulting behavior of \(\alpha_\omega\) is shown in Fig.~\ref{alpha_vs_lnt.pdf}. Over the accessible critical region, the exponent remains close to zero and exhibits a stable plateau. Within the accessible near-critical region, the effective logarithmic slope approaches zero, indicating the absence of a positive-power divergence of \(C_J\). Since a vanishing local logarithmic slope may also arise from a finite regular background, we further examine the one-sided behavior of \(C_J\) across the CEP. The finite limiting values are consistent with the mean-field expectation \(\alpha_\omega=0\).

\begin{figure}[htbp]
\centering
\includegraphics[width=0.48\textwidth]{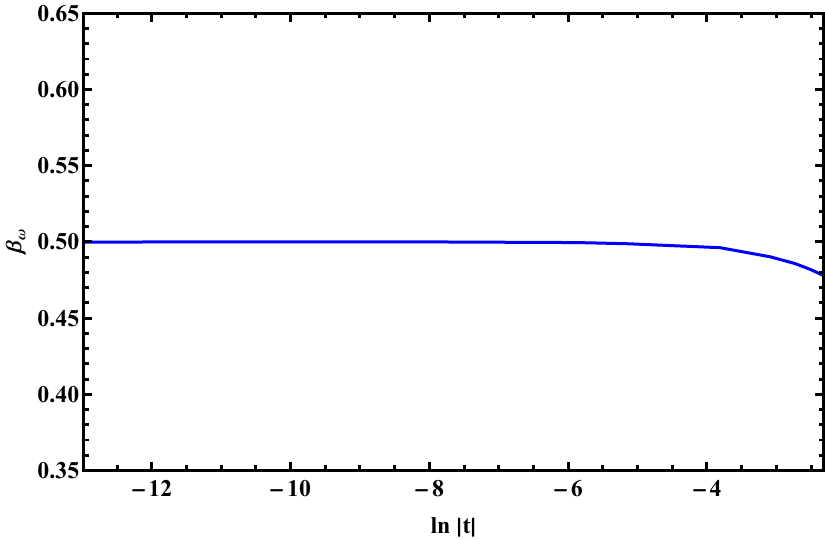}
\caption{
(Color online) Effective critical exponent \(\beta_\omega\) as a function of \(\ln|t|\), extracted from the discontinuity of the rotational polarization along the first-order transition line.
}
\label{fig:beta}
\end{figure}

We next investigate the critical exponent \(\beta_\omega\), which characterizes the vanishing of the discontinuity of the order-parameter-like quantity when the CEP is approached along the first-order transition line. In the rotating system, the thermodynamic quantity conjugate to the angular velocity is the rotational polarization,
\(
J=-\frac{\partial\Omega}{\partial\omega}.
\)
Although \(J\) is not the conventional chiral order parameter, its singular part couples to the soft chiral mode through the self-consistent constituent mass. Provided that \(\Omega_{\omega M}^{\rm CEP}\neq0\), this singular contribution inherits the leading critical scaling near the CEP.
Along the first-order transition line, two thermodynamically degenerate phases coexist at the same \((T,\omega)\), corresponding to two different solutions of the gap equation with constituent masses \(M_>\) and \(M_<\). The corresponding rotational polarizations are denoted by \(J_>\) and \(J_<\), respectively, and the discontinuity is defined as
\(
\Delta J=J_>-J_< .
\)
As the CEP is approached from the first-order side, the two phases become indistinguishable and the discontinuity gradually vanishes. The corresponding scaling behavior is described by
\(
|\Delta J|\sim |t|^{\beta_\omega}.
\)
 As shown in Fig.~\ref{fig:beta}, \(\beta_\omega\) exhibits a broad plateau in the near-critical region and remains close to
\(
\beta_\omega\simeq \frac12.
\)
The small deviation observed away from the CEP originates from non-asymptotic effects, where the system gradually leaves the critical scaling regime. The numerical result is therefore consistent with the mean-field value associated with the leading order-parameter scaling.

\begin{figure}[htbp]
\centering
\includegraphics[width=0.48\textwidth]{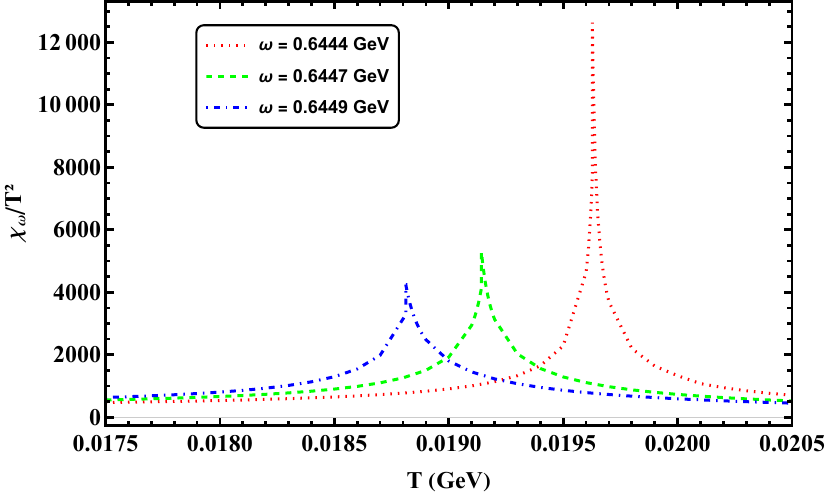}
\caption{
(Color online) Scaled rotational susceptibility \(\chi_\omega/T^2\) as a function of temperature for several angular velocities around the CEP.
}
\label{fig:chi-T}
\end{figure}

We then investigate the critical behavior of the rotational susceptibility,
\(
\chi_\omega=(\frac{d J}{d\omega})_T,
\)
which characterizes the response of rotating QCD matter to variations in angular velocity. Fig.~\ref{fig:chi-T} shows the temperature dependence of dimensionless scaled susceptibility \(\chi_\omega/T^2\) for several angular velocities around \(\omega_{\mathrm{CEP}}\). Away from the critical region, the susceptibility exhibits only a moderate enhancement and varies smoothly across the transition region. As \(\omega\) is decreased from \(0.6449\) to \(0.6444\) GeV, the fixed-\(\omega\) trajectory moves progressively closer to \(\omega_{\rm CEP}\). As the CEP is approached, the response becomes increasingly enhanced, accompanied by a sharper peak structure near the critical temperature. This behavior indicates the strong amplification of the rotational response associated with the softening of the critical mode near the CEP.

\begin{figure}[htbp]
\centering
\includegraphics[width=0.48\textwidth]{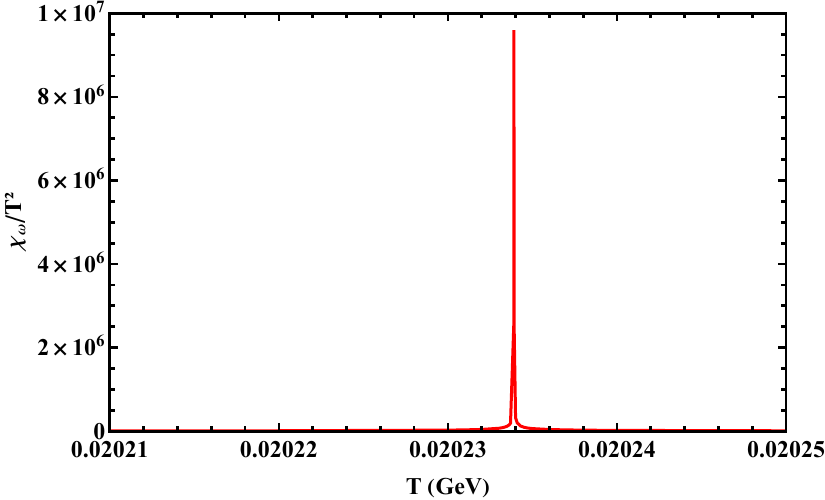}
\caption{
(Color online)  Scaled rotational susceptibility \(\chi_\omega/T^2\) as a function of temperature at
\(\omega=\omega_{\mathrm{CEP}}\).
}
\label{fig:chi-CEP}
\end{figure}

\begin{figure}[htbp]
\centering
\includegraphics[width=0.48\textwidth]{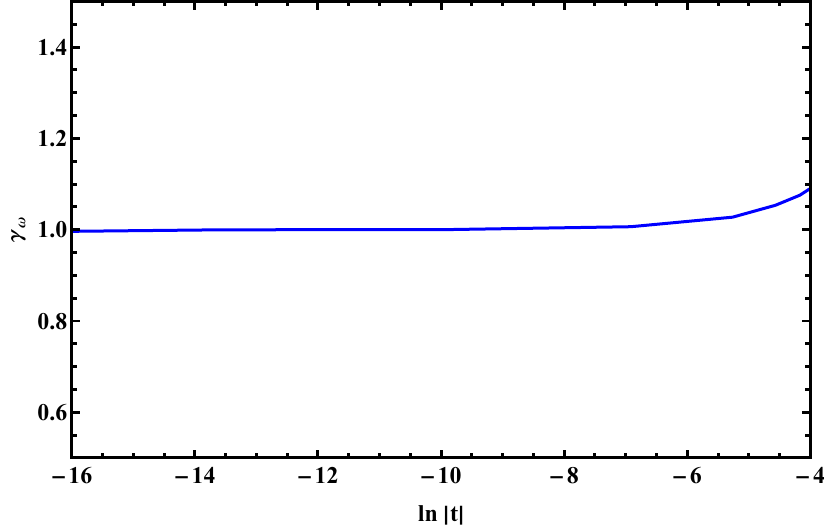}
\caption{
(Color online) Effective critical exponent \(\gamma_\omega\) as a function of \(\ln|t|\),  along a trajectory tangent to the phase boundary near the CEP.
}
\label{fig:gamma}
\end{figure}

At the critical angular velocity \(\omega=\omega_{\mathrm{CEP}}\), the temperature dependence of  dimensionless scaled susceptibility \(\chi_\omega/T^2\)  is presented in Fig.~\ref{fig:chi-CEP}. A sharply enhanced peak develops near  \(T\simeq T_{\mathrm{CEP}}\), signaling the singular growth of the rotational response. Within the mean-field framework, this behavior originates from the vanishing curvature of the thermodynamic potential in the chiral direction,
\(
\frac{\partial^2\Omega}{\partial M^2}\rightarrow0,
\)
which enhances the coupling of the rotational response to the soft chiral mode and leads to a divergent susceptibility. Fig.~\ref{fig:chi-CEP} is intended to illustrate this characteristic enhancement in the vicinity of the CEP rather than to provide the data used for extracting the critical exponent.
Since \(\chi_\omega=\chi_\omega(T,\omega)\), the CEP can be approached along different thermodynamic trajectories in the \((T,\omega)\) plane. For the quantitative determination of \(\gamma_\omega\), we approach the CEP along a trajectory tangent to the phase boundary at the CEP, in analogy with the conventional prescription employed in studies of critical susceptibility in the \(T-\mu\) plane. Along this trajectory, the susceptibility obeys
\(\chi_\omega\sim |t|^{-\gamma_\omega}\).
The corresponding effective exponent is shown in Fig.~\ref{fig:gamma}. As the CEP is approached, \(\gamma_\omega\) develops a stable plateau close to 1. The deviation from the asymptotic value at larger distances from the CEP reflects preasymptotic effects due to the departure from the critical scaling regime.

\begin{figure}[htbp]
\centering
\includegraphics[width=0.48\textwidth]{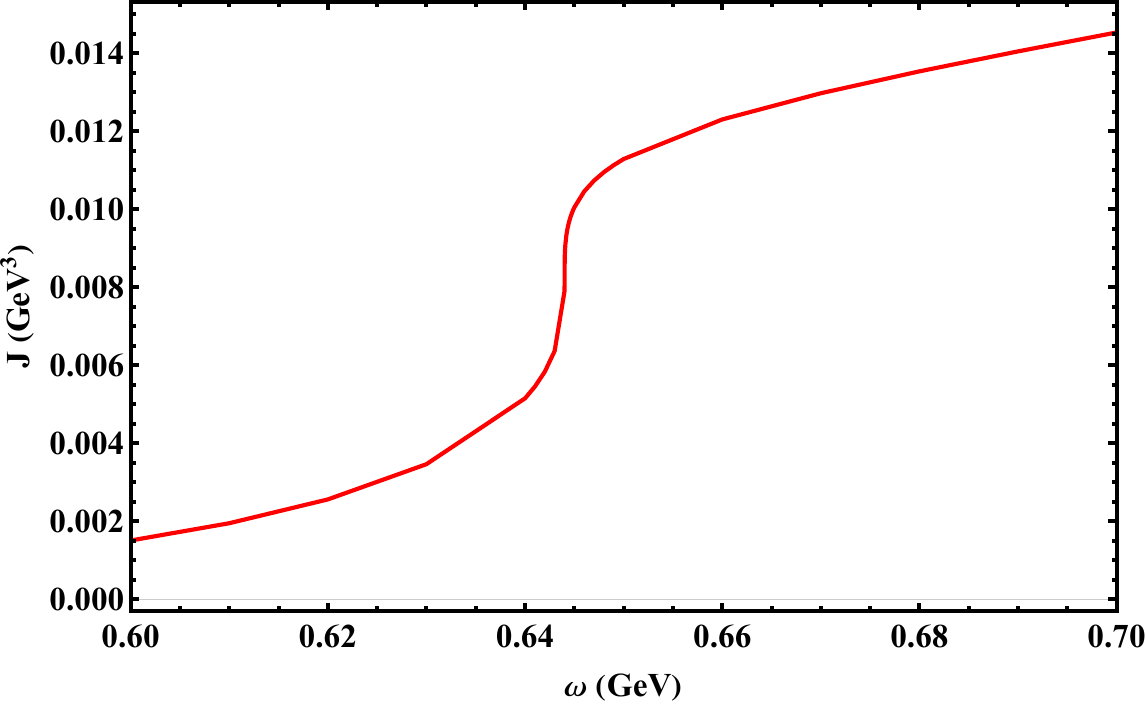}
\caption{(Color online)
Rotational polarization \(J\) as a function of angular velocity \(\omega\) at \(T=T_{\mathrm{CEP}}\). }
\label{J-omega}
\end{figure}

\begin{figure}[htbp]
\centering
\includegraphics[width=0.48\textwidth]{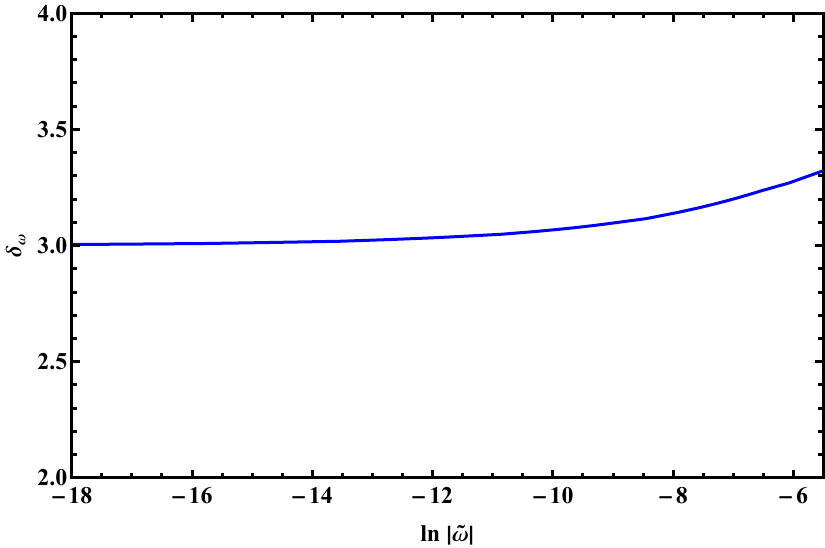}
\caption{(Color online) Effective critical exponent \(\delta_\omega\) as a function of \(\ln |\tilde{\omega}|\), along the critical isotherm \(T=T_{\mathrm{CEP}}\). }
\label{fig:delta}
\end{figure}

Figs.~\ref{J-omega} and \ref{fig:delta} provide complementary evidence for the critical-isotherm scaling behavior of rotating QCD matter near the CEP. As shown in Fig.~\ref{J-omega}, the rotational polarization \(J\) exhibits a nonlinear dependence on the angular velocity \(\omega\) along the critical isotherm \(T=T_{\mathrm{CEP}}\), with a rapid variation in the response near \(\omega_{\mathrm{CEP}}\). This behavior reflects the enhanced nonlinear response associated with criticality and suggests that the angular velocity provides an operational field-like control direction in the vicinity of the CEP. To quantify this critical response, we extract the effective exponent \(\delta_\omega\) according to the scaling relation
\(
\tilde{J}\sim |\tilde{\omega}|^{1/\delta_\omega}.
\)
The exponent is determined exclusively from the near-critical local-slope analysis shown in Fig.~\ref{fig:delta}. As \(|\tilde{\omega}|\rightarrow0\), \(\delta_\omega\) approaches the mean-field value
\(
\delta_\omega\simeq3,
\)
indicating that the asymptotic critical-isotherm behavior is consistent with the Landau mean-field prediction. The deviation from the asymptotic value at larger \(|\tilde{\omega}|\) can be attributed to preasymptotic effects and nonsingular background contributions outside the scaling regime.

Together with the other effective critical exponents, these results demonstrate that angular velocity provides an additional thermodynamic control direction for probing CEP criticality in the \((T,\omega)\) plane, while the extracted effective exponents remain consistent with the Landau mean-field scaling pattern.

\section{CONCLUSIONS\label{sec4}}
The present study establishes a thermodynamic scaling framework for a rotation-induced critical endpoint in the two-flavor NJL model. Beyond determining the location of the CEP in the \((T,\omega)\) plane, we demonstrate that the rotational degree of freedom can be incorporated into the critical thermodynamic description through the conjugate pair $(\omega,J)$. Although the rotational polarization $J$ is not the conventional chiral order parameter, its singular contribution couples to the critical mode associated with the CEP through the mixed dependence of the thermodynamic potential on the rotational and chiral sectors. As a consequence of the nonvanishing mixed coupling between the rotational and chiral sectors, the singular parts of the rotational response functions inherit the critical behavior associated with the soft chiral mode. In particular, the rotational susceptibility is strongly enhanced near the CEP due to the vanishing curvature $\Omega_{MM}$ in the critical direction.

Using different thermodynamic trajectories approaching the CEP, we extract the effective critical exponents associated with the specific heat density, coexistence discontinuity, rotational susceptibility, and critical-isotherm response of the rotational polarization. The resulting effective critical exponents,
\(\alpha_\omega\simeq0, 
\beta_\omega\simeq\frac{1}{2}, \gamma_\omega\simeq1,
\delta_\omega\simeq3\), satisfy the standard scaling relations \(\alpha_\omega+2\beta_\omega+\gamma_\omega=2,
\alpha_\omega+\beta_\omega(1+\delta_\omega)=2\), and exhibit the characteristic Landau mean-field behavior within the present framework. These results indicate that angular velocity provides an additional thermodynamic control direction in which CEP criticality can be probed, while the extracted effective exponents remain consistent with Landau mean-field scaling. At the same time, the rotational observables provide an alternative manifestation of CEP criticality through the thermodynamic channel associated with angular momentum. A quantitative investigation beyond mean-field approximation, including fluctuation effects and a more realistic treatment of inhomogeneous rotating matter, will be necessary to determine how the present scaling behavior is modified in realistic vortical QCD systems.

Several extensions are warranted. Finite-volume effects and boundary conditions should be incorporated to assess the robustness of the present local analysis~\cite{Braun2005,Klein2017,Ebihara:2016fwa,Xu2020,Correa:2023nmf,Chen:2022mhf}, while additional interaction channels  such as vector interactions~\cite{Wang:2018sur}, and the Polyakov--Nambu--Jona-Lasinio (PNJL) model~\cite{Meisinger:1995ih,Meisinger:2001cq,Fukushima:2003fw,Mocsy:2003qw,Megias:2004hj,Ratti:2005jh,Fukushima:2008wg} may modify the phase structure and CEP location. More importantly, the present calculation is restricted to mean field and therefore does not include critical fluctuations beyond mean field. Approaches such as the functional renormalization group~\cite{Chen:2023cjt} will be required to determine whether and how the effective scaling behavior obtained here is modified by fluctuations and to establish the universality properties of a rotation-induced CEP beyond the present approximation.
\\

\section*{Acknowledgments}
We would like to thank Wei-jie Fu, Fei Gao, Gaoqing Cao, Xin-Li Sheng, Hao-Lei Chen, Rui Wen, and Zhibin Li for useful discussions. The work was supported by the National Natural Science Foundation of China (NSFC) (12375137, 12405154, 12575144), and the Yichang Natural Science Foundation (A25-3-002).

\bibliography{ref-lib}
\end{document}